\documentstyle[preprint,aps,epsfig]{revtex}

	 \def\be{\begin{equation}}
	 \def\ee{\end{equation}}
	 
	 \def\be{\begin{equation}}
	 \def\bea{\begin{eqnarray}}
	 \def\t{\tau}
	 \def\o{\over}
	 
	 \def\a{\alpha}
	 \def\b{\beta}
	 \def\e{\eta}
	 \def\ee{\end{equation}}
	 \def\eea{\end{eqnarray}}
	 \def\R{\rm {I\kern-.200em R}}
	 \def\C{\rm {I\kern-.520em C}}
	 \def\c{\chi}
	 \def\s{\sigma }
	 \def\A{{\hat A}}
	 \def\D{{\hat D}}
	 \def\E{{\hat E}}
	 \def\at{\tilde {\alpha}}
	 \def\bt{\tilde {\beta}}

	 \def\y{Y(N,M)}
	 \def\yd{Y_D(N,M)}
	 \def\c{\sum_{\rm conf}}
	 \def\ot{\otimes}

\begin{document}
\begin{titlepage}

\begin{center} {\large \bf An Exactly Solvable Two-Way Traffic Model with  
Ordered Sequential Update } \\

\vskip 1cm

\centerline {\bf M.E. Fouladvand$ ^{a,c}$ \footnote
{e-mail: foolad@netware2.ipm.ac.ir},
H.-W. Lee$ ^{b}$ \footnote {e-mail: hwlee@ctp.snu.ac.kr}} 
\vskip 1cm
{\it $^a$ Department of Physics, Sharif University of Technology, }\\
{\it P.O.Box 11365-9161, Tehran, Iran }\\
{\it $^b$ Center for Theoretical Physics, Seoul National University,}\\
{\it Seoul, 151-742, Korea}\\
{\it $^c$ Institute for Studies in Theoretical Physics and Mathematics,}\\
{\it P.O.Box 19395-5531, Tehran, Iran}
\end{center}

\begin{abstract}

Within the formalism of the matrix product ansatz, we study a two-species
asymmetric exclusion process with backward and forward site-ordered
sequential updates.  This model, which 
was originally introduced with the random sequential update \cite{lee}, 
describes a two-way traffic flow with a {\it dynamic impurity } and
shows a phase transition between the free flow and the traffic jam.
We investigate characteristics of this
jamming and examine similarities and differences between 
our results and those with the random sequential update. 

\end{abstract}

\vspace*{0.5cm} 


\end{titlepage}

\vspace {1 cm}
\newpage
\section{Introduction}

The one dimensional asymmetric simple exclusion process (ASEP) has been the 
subject of rigorous and intensive studies in recent years \cite{ziabook,book}. 
A variety of phenomena can be modeled by the ASEP and 
its generalizations (see \cite{book,schutzbook} and references therein). 
The model has a natural interpretation
as a description of traffic flow (\cite{wolf} and references therein) 
and constitutes a basis for more realistic ones \cite{nagel}. 
In traffic flow theories, the formation of traffic jams 
is one of the fundamental problems. 
Apart from their spontaneous formation \cite{nagel}, 
they can be also produced by hindrances such as road works or slow cars. 
Although these hindrances act locally, 
they may induce global and macroscopic effects on a system. This 
kind of behavior is one of the characteristic properties in non-equilibrium 
systems and has been studied in the context of driven lattice gases 
\cite{kandel,janowsky,derida2asep,schutz1,mallick,lee,kolom}. 

In the ASEP models, two kinds of impurities are discussed 
in the literature. The first one is ``{\it dynamic impurities}'',
i.e., defective particles which jump with a rate  lower
than others~\cite{derida2asep,mallick,lee,evans}. In the 
traffic terminology, such moving defects can be visualized 
as slow cars on a road, which in certain situations can 
induce a phase transition from the free flow to the congested flow.
The other kind of impurities is ``{\it static impurities}'' 
such as imperfect links where the hopping rate is 
lower than in other links~\cite{janowsky,schutz1,kolom,barma,emmerich,yukawa}. 
Static impurities can also produce shocks 
in a system~\cite{janowsky,schutz1}. 
For either type of impurities, a limited amount of
exact results is available and most of them are for models
with the random sequential update~\cite{derida2asep,mallick,lee,barma}.
For the fully parallel update, which is most realistic 
in traffic flow problems, exact solution are very 
rare~\cite{nagel,evans,rajdegier,niehuis} and most studies instead utilize 
approximation methods or 
numerical approaches~\cite{yukawa,csahok,knopse}. 

Recently a two-way traffic model is introduced \cite{lee} where
cars move forward in one lane and trucks move backward in the other lane, 
and both cars and trucks reduce their speeds  when they approach each other. 
Within the matrix product ansatz (MPA) formalism, 
a modified version of this model is solved exactly
in the particular case when there is only one truck in the system. 
This truck behaves as a dynamic impurity  and induces a phase transition
between the free flow and the congested flow of the cars. 
Various characteristics of the phase transition are 
examined.

In this paper we study an exactly solvable traffic model with two
types of ordered sequential updates, which is identical to
the model studied in Ref.~\cite{lee} except for the updating schemes.
More precisely, the updating schemes that we consider are the backward
and forward site-ordered sequential updating schemes 
in which one updates links of the chain sequentially. 
Alternatively one may use the so-called 
particle-ordered sequential updating scheme 
in which one sequentially updates the positions 
of the particles~\cite{evans}. In this paper we restrict ourself
to the site-ordered sequential updates.

Presently it is of prime interest to determine whether distinct updating
schemes can produce different behaviors. 
The implementation of the type of update is an essential part 
of the definition of a model and some characteristics
of the model may change dramatically.
The aim of this work is to investigate consequences of 
changing the updating scheme of the model. Then combined
with the results in the random sequential updating scheme~\cite{lee},
we examine similarities and differences between 
the results in different types of update. 
Our approach utilizes a mapping between the quadratic algebras
of the ordered and random sequential updates, which has been initiated 
in the context of the one-species ASEP \cite{raj96,raj98}.

This paper has the following organization.
In section II we define the model, construct our MPA with both 
backward and forward sequential updates,
and obtain the relevant quadratic algebras. 
Section III presents the expressions of the average velocities 
of cars and truck, their thermodynamic limits, and 
a comparison with the corresponding results in Ref.~\cite{lee}.
In section IV, we consider density profiles of cars and compute
the probability to find a car at a distance $x$ from the truck, 
which in the high density phase appears as a shock. 
We discuss the ambiguity in specifying the density profile,
which is related with the nature of the updates, and introduce
a new definition of the averages to avoid the ambiguity.
Section V is devoted to the density-density correlation function.
In section VI, we study the model in the presence of two trucks and evaluate
the probability of their distance being $R$ . The paper ends with some
concluding remarks in section VII.

\section{Model Definitions and Matrix Product Ansatz}

\subsection{Two-way traffic model with random sequential update}

Here we  describe briefly the two-way traffic model
with random sequential update (RSU) introduced in \cite{lee}. 
Consider two parallel one dimensional chains, 
each with $N$ sites. The periodic boundary condition applies
to each chain. There are $M$ cars and $K$ trucks in the first 
and the second chain, respectively. Cars move to the right 
and trucks move to the left. The state of the 
system is characterized by two sets of occupation numbers 
$(\t_1,\t_2,\cdots,\t_N)$ and $(\sigma_1,\sigma_2,\cdots,\sigma_N )$ 
 for the first and the second lanes. 
If the site $i$ of the car lane is occupied by a car, 
$\t_i =1$ and zero otherwise. Similarly $\sigma_i =1$
if the site $i$ of the truck lane is occupied by a truck 
and $\sigma_i =0$ if the site is empty. 
In an infinitesimal time interval $dt$, 
a car (truck) hops to its right (left)
empty site with the probability $dt$ ($\gamma dt$) 
if there is no truck (car) in front, and
with the probability reduced by a factor $\beta$ otherwise.
More explicitly one has
\bea
(\t_i,\t_{i+1})=(1,0) \rightarrow (0,1) \; {\rm with} \ \ {\rm rate} 
  \ \ \left\{  \begin{array}{cl} 
\displaystyle 1 &  {\rm if} \ \  \sigma_{i+1}=0 \\
\displaystyle \rule{0mm}{7mm} {1\o {\b}} &  {\rm if} \ \  \sigma_{i+1}=1 \ \  
  ({\rm truck} \ \ {\rm in}  \ \ {\rm front})   
\end{array} \right.
\eea
\bea
(\s_i,\s_{i+1})=(0,1) \rightarrow (1,0) \; {\rm with} \ \ {\rm rate}  
\ \ \left\{ \begin{array}{ll} 
\displaystyle \gamma & {\rm if} \ \ \t_{i}=0 \\
\displaystyle \rule{0mm}{7mm} {\gamma \o \b } & {\rm if} \ \  \t_{i}=1 \ \ 
 ( {\rm car} \ \ {\rm in}  \ \ {\rm front}) \ .   
\end{array} \right. 
\eea
The reduction factor $\b$, which ranges between one and infinity,
is related to the width of roads:
$\b =1$ corresponds to a very wide 
road or a highway with a lane divider and $\b =\infty $ 
corresponds to a one lane road. 
Simulations with finite densities of cars and trucks show \cite{lee} 
that in the steady state, the average velocities of cars and trucks
decrease smoothly with increasing $\beta$.
As an interesting limiting case, situations with a single truck 
is considered while the density of cars is kept finite. 
For this particular case, simulations suggest that
for a given density $n$ of cars, 
there exists a density-dependent critical value $\b_c$,
below which the average velocity of cars remains constant 
and above which the average velocity decreases linearly 
with respect to $1- {1\o {\b}}$, a measure of the road narrowness.
For $\beta>\beta_c$, the simulation also finds 
the phase segregation into high (traffic jam)  
and low (free flow) density regions.

To investigate the characteristics of this single truck case analytically, 
the above two-lane model has been  modified to an exactly solvable one. 
If one forbids a car and a truck to occupy two parallel site $i$ 
simultaneously, one can describe configurations
with a single set of occupation numbers $\{\t_i\}$ 
where $\t_i = 0$ (empty site), $1$ (occupied by a car), or 
$2$ (occupied by a truck).
The following rules describe the modified dynamics:
\begin{eqnarray}
(1,0) \rightarrow (0,1) \; \ \ \ & {\rm with}\ \  {\rm rate} \ \ & 1
 \nonumber \\
(0,2) \rightarrow (2,0) \; \ \ \ & {\rm with}\ \  {\rm rate} \ \ & 
   \gamma 
 \\
(1,2) \rightarrow (2,1) \; \ \ \ & {\rm with}\ \  {\rm rate} \ \ & 
 {1\o {\b}} \nonumber \ .
\end{eqnarray}

This model is equivalent to a two-species ASEP and can be 
solved exactly by the method of the matrix product state (MPS).
The steady state weight $P_s$ of a given
configuration $(\t_1,\t_2,\cdots,\t_N)$ is proportional to 
the trace of the normal product of some matrices:
\be
P_s (\t_1,\t_2,\cdots,\t_N) \sim {\rm Tr}(X_1X_2\cdots X_N)
\label{MPSforRSU}
\ee
where
\bea
X_i = \ \ \left\{ \begin{array}{cc}   

D  & {\rm for} \ \t_i=1 \\
E  & {\rm for} \ \t_i=2 \\
A  & {\rm for} \ \t_i=0 \\

\end{array} \right.
\eea
and these matrices satisfy the quadratic algebra 
\be
DE= D+E , \ \  \a AE=A , \ \  \b DA= A  \ \ (\a\equiv \b\gamma) \ .
\label{for1}
\ee
With the help of the MPS method, many important characteristics 
such as average velocities, density profiles, 
 and $k$ point correlation functions have been obtained exactly 
in Ref.~\cite{lee}.

\subsection{Two-way traffic model with ordered sequential updates}

In the ordered sequential updating (OSU) schemes,
time is discrete and the update occurs in an ordered way.
To describe the dynamics in this scheme, it is convenient to introduce 
a Hilbert space that is spanned by the orthonormal ket
vectors $|\{\tau\}\rangle=|\tau_1\rangle \ot |\tau_2\rangle 
\ot \cdots \ot |\tau_N\rangle$.
The state of a system at the $j$-th time step can be 
represented by a ket vector $|P,j\rangle$
\be
|P,j\rangle=\c P(\{\tau\};j)|\{\tau\}\rangle
\ee
where $P(\{\tau\};j)$ is the weight of the configuration
$\{\tau\}$ at the $j$-th time step and $\c$
denotes the summation over all possible configurations.
The state at the next time step is then obtained
from  $|P,j\rangle$  by applying a transfer matrix 
\be
|P,j+1\rangle= T|P,j\rangle 
\ee
where the transfer matrix $T$ takes a different form
depending on the precise nature of the updates. 
In the backward sequential updating (BSU) scheme, for example, 
the transfer matrix $T$ becomes $T_{\leftarrow}$ where
\be
T_{\leftarrow} = T_{N,1}T_{1,2} \cdots T_{N-2,N-1}T_{N-1,N}
\label{for2}
\ee
and in the forward sequential updating (FSU) scheme, 
it becomes $T_{\rightarrow}$ where
\be
T_{\rightarrow} = T_{N,N-1}T_{N-1,N-2} \cdots T_{2,1}T_{1,N} \ .
\label{FSUtransfer}
\ee 
Each element in the products is defined by 
\be
T_{i,i+1}=T_{i+1,i}=
  \underbrace{{\bf 1}\ot\cdots \ot {\bf 1}}_{i-1} \ot {\cal T}\ot
  \underbrace{{\bf 1}\ot\cdots \ot {\bf 1}}_{N-i-1} \ .
\label{localtransfer}
\ee
Here ${\bf 1}$ is the identity matrix acting on
a local ket vector $|\tau\rangle$ and 
the local transfer matrix ${\cal T}$ acts on the tensor product state of
two local ket vectors, $|\tau\rangle \ot |\tau'\rangle$.
In Eqs.~(\ref{for2},\ref{FSUtransfer}), the site $N$ is chosen
as a starting point of the update for definiteness. 

The local transfer matrix ${\cal T}$ varies depending on
possible exchange processes allowed in a model. 
As a straightforward generalization of
the two-way traffic model in Ref.~\cite{lee}, we allow
the following processes to occur in each discrete time step:
\begin{eqnarray}
(1,0) \rightarrow (0,1) \; \ \ & {\rm with} \ \  {\rm probability} \ \ & 
  \e   \nonumber \\
(0,2) \rightarrow (2,0) \; \ \ & {\rm with} \ \  {\rm probability} \ \ & 
  \e \gamma  \\
(1,2) \rightarrow (2,1) \; \ \ & {\rm with} \ \  {\rm probability} \ \ &
   {\e \over \b} \rule{0mm}{6mm} \nonumber \  
\end{eqnarray}
where $0 \le \e,\e\gamma,{ \e \over \b } \le 1 \le \b$.
For this two-way traffic model, off-diagonal matrix elements of
${\cal T}$ have nonzero values only for the following elements
\begin{eqnarray}
\left( \langle 0|\ot \langle 1|\right) {\cal T} 
  \left( |1 \rangle \ot |0 \rangle \right) &=& \e
  \nonumber \\
\left( \langle 2|\ot \langle 0|\right) {\cal T} 
  \left( |0 \rangle \ot |2 \rangle \right) &=& \e \gamma
  \\
\left( \langle 2|\ot \langle 1|\right) {\cal T} 
  \left( |1 \rangle \ot |2 \rangle \right) &=&  {\e \over \b}
  \rule{0mm}{6mm} \nonumber 
\end{eqnarray}
and the diagonal elements can be specified from
the condition of the probability conservation, 
$\sum_{\tau_3,\tau_4}
\left( \langle \tau_3|\ot \langle \tau_4|\right)  {\cal T} 
\left( |\tau_1 \rangle \ot |\tau_2 \rangle \right)=1$, 
which ensures that the sum of entries in each column is equal to one.

\subsection{Matrix product state} 

Recently Krebs and Sandow \cite{sandkrebs} showed that
the steady state of any stochastic process with 
arbitrary nearest neighbor interactions, defined on
an open one dimensional systems with RSU, can be always written 
as a matrix product state. 
This theorem was soon generalized by 
Rajewsky and Schreckenberg \cite{rajsch} to stochastic processes 
with ordered sequential and sub-lattice parallel updates. 
For closed systems with the periodic boundary condition,
on the other hand, such a theorem is not available at present
and it is not yet clear how widely the MPA is applicable.
Nevertheless, numerous recent 
studies~\cite{derida2asep,mallick,lee,raj98,hinsan,Arndt,jafar} 
have  proven  that
the MPA can be a powerful investigation tool even for
closed systems as well. Motivated by recent successes, we here assume
that the stationary state of the two-way traffic model
can be written in terms of the MPS:
\be
P_s(\t_1,\t_2,\cdots,\t_N) \sim {\rm Tr}(X_1X_2\cdots {\hat X}_N) \ 
\label{for3}
\ee
where $X_i=D,E,$ or $A$ ($\hat{X}_N=\hat{D},\hat{E},$ or $\hat{A}$)
depending on $\t_i$ ($\t_N$).
Note that the matrices at the site $N$ are different from those
at other sites\footnote{
The hatted-matrix in Eq.~(\ref{for3}) breaks 
the translational invariance of the problem. As a result,
the steady state weights are not completely fixed 
by the relative distances between cars and trucks. 
This situation is in contrast to Ref.~\cite{lee} 
where the weights depend on the relative distances only
due to the translational invariance [see~(\ref{MPSforRSU})].}, 
which stems from the special
role of the site $N$ as a starting point of the update.
The necessity of the hatted matrices in the OSU scheme is clearly
explained in Ref.~\cite{raj98}.

By introducing two ket vectors $|U\rangle$ and $|\hat{U}\rangle$,
\begin{eqnarray}
|U\rangle &=& A|0\rangle+D|1\rangle+E|2\rangle \nonumber \\
|\hat {U} \rangle &=& \hat{A}|0\rangle+\hat{D}|1\rangle+\hat{E}|2\rangle \ ,
\end{eqnarray}
where the coefficients $A,D,E,\A,\D$, and $\E$ are
matrices defined on a different auxiliary space, 
we can write (\ref{for3}) formally as 
\be
\vert P_s \rangle= {1\o {Z(N,M)}} \ 
  {\rm Tr}\left(|U\rangle \ot |U\rangle 
    \ot \cdots \ot|{\hat U}\rangle \right) 
\label{for4}
\ee 
in which  the normalization constant $Z(N,M)$ is the sum 
of all weights
\be
Z(N,M)=\c {\rm Tr}( X_1\cdots {\hat X_N})
\label{probsum}
\ee
and the trace is taken only over the normal product of
the matrices $A,D,E,\hat{A},\hat{D},\hat{E}$.

By definition, $|P_s\rangle$ should be stationary
under the action of the transfer matrix $T$
$$
T|P_s\rangle=|P_s\rangle \ .
$$ 
In the BSU scheme, the stationarity  is guaranteed if 
the following relation holds:
\be
{\cal T}(|U\rangle \ot|{\hat U}\rangle )=
  |{\hat U}\rangle \ot |U\rangle \ 
\label{for5}
\ee
which simply implies that upon the action of $T_{\leftarrow}$,
the ``{\it defect}'' $|{\hat U}\rangle$ is transferred 
backward through the chain and returns back to 
the site $N$, its initial position. 
Eq.~(\ref{for5}) then leads to the following quadratic algebra:
\begin{eqnarray}
 & [A,\A]=0,\ \ [D,\D]=0,\ \ [E,\E]=0 &
  \nonumber \\
& A\D+\e D\A = \A D , \ \ (1-\e)D\A=\D A &
  \nonumber \\
& \displaystyle 
  \e \gamma A\E + E\A=\E A, \ \ (1-\e \gamma )A \E = \A E &
\label{for6}
  \\
& \displaystyle
  {\e \over \b} D\E +E\D=\E D, \ \ (1-{\e \over \b})D\E= \D E & \ 
  \nonumber
\end{eqnarray}

In the FSU scheme, on the other hand, the stationarity
requires
\be
{\cal T}(|\hat{U}\rangle \ot| U\rangle )=
  |U\rangle \ot |\hat{U}\rangle \ 
\label{stationaryFSU}
\ee
which implies that the ``{\it defect}'' $|\hat{U}\rangle$
is now transferred in the forward direction.
Eq.~(\ref{stationaryFSU}) leads to a quadratic algebra, which is
identical to (\ref{for6}) upon the replacements
$A \leftrightarrow \A, D \leftrightarrow \D, E \leftrightarrow \E$.

\subsection{Mapping onto the RSU algebra} 

In order to calculate physical quantities using the MPS, 
it is in principle necessary to find a representation of the matrices
that satisfies the relevant quadratic algebras for the BSU and
FSU schemes. In some cases, however, this job can be avoided and 
the quadratic algebra itself is sufficient for calculations 
\cite{derida2asep,lee,evans,rajdegier,niehuis,vahid1} .
 Also recent studies on ASEP~\cite{raj96,raj98} and
multi-species ASEP~\cite{vahid1,foulad,vahid2} have
demonstrated that the algebra for a stochastic process
in an OSU scheme can be mapped onto the algebra for a related stochastic
process in the RSU scheme, for which an explicit representation
of the matrices are known. Here we show that 
this mapping holds for the two-way traffic model as well.
We first assume 
\be
\A=A+a, \ \ \D=D+d,\ \ \E=E +e
\label{for8}
\ee
where $a,d$ and $e$ are real numbers.
In the BSU scheme, it can be verified that
the following choice 
\be
a=0,\ \ d=-{\e \over \b}, \ \ e={\e \o {\b - \e} }  
\ee                                    
maps the algebra (\ref{for6}) to 
\be
DE=D+E, \ \ \at AE= A , \ \ \bt DA=A \ 
\label{algebramapping}
\ee
where
\be
\at={\a( \b - \e) \o \b -\a \e }  ,\ \  
\bt =\b \ .
\ee
Note that the algebra (23) is identical to the RSU algebra~(\ref{for1})
except for the renormalization of $\a$ to $\at$.

In the FSU scheme, one may use the mapping
to the BSU algebra by simply interchanging
the roles of $A,D,E$ with $\hat{A},\hat{D},\hat{E}$ 
[see Eqs.~(\ref{for5},\ref{stationaryFSU})].
However it turns out that it is more convenient 
for the subsequent analysis to use a direct mapping to the RSU scheme.
We again assume Eq.~(\ref{for8}) and choose
\be
a=0,\ \ d={\e \over \b-\e }, \ \ e=- {\e \over \b} \ .  
\label{adeFSU}
\ee                                    
Then the FSU algebra is mapped to Eq.~(\ref{algebramapping})
with
\be
\at=\a  ,\ \  
\bt ={\b - \e \over 1-\e} \ .
\label{atbtFSU}
\ee
Note that now $\b$ is renormalized to $\bt$.

\section{Average Velocities}

In this section  we consider the special case where there is only a single
truck in the system and evaluate the average velocities of cars 
and the truck.

\subsection{BSU scheme}
 
The movement of cars from a site, $k$ for example, to
its neighboring site $k+1$ is achieved by $T_{k,k+1}$.
Then the average current of cars $\langle J^{\rm car}_k \rangle$
between the two sites can be written as follows:
\be
\langle J_k^{\rm car}\rangle= {1\o {Z(N,M)}}
\left\{ \e \c^{N,M} {\rm Tr} (\underbrace{\cdots}_{k-1\ {\rm sites} }  
 D \A  \cdots) 
+ {\e \over \b} \c^{N,M} {\rm Tr} 
  (\underbrace{\cdots}_{k-1\ {\rm sites}}  D \E \cdots) \right \} \      
\label{CARCURRENT}
\ee 
where the specifications  $N$ and $M$ in the summations
represent that the matrix products $(\cdots D \A \cdots)$ and
$(\cdots D \E \cdots)$ contain $N$ matrices and $M$ of them
denote cars. Below, these specifications will be omitted 
in usual situations, that is, when they are $N$ and $M$.
In Eq.~(\ref{CARCURRENT}), it has been taken into account that
before the action of $T_{k,k+1}$, the steady state
is acted by $T_{k+1,k+2} \cdots T_{N-1,N}$,
which shifts the hat from the site $N$ to $k+1$. 
More rigorous derivation is given in Appendix A. 
In a similar way, the average current of a truck
$\langle J^{\rm truck}_{k} \rangle$ between 
the site $k$ and $k+1$ reads
\be
\langle J_k^{\rm truck}\rangle= {1\o {Z(N,M)}}
\left\{ \e \gamma \c {\rm Tr} 
  (\underbrace{\cdots}_{k-1}  A \E  \cdots) 
+ {\e \over \b} \c {\rm Tr} 
  (\underbrace{\cdots}_{k-1} D \E  \cdots) \right \} \ .     
\label{truckcurrent}
\ee 

It is instructive to briefly sketch the calculation 
of Eq.~(\ref{CARCURRENT}).
Using the cyclic invariance and the fact that 
in the first term of Eq.~(\ref{CARCURRENT}), 
the single truck can locate anywhere on the remaining $N-2$ sites, 
we find
\be
\langle J_k^{\rm car}\rangle = {1\o {Z(N,M)}}
\left \{\e \sum _{i=1}^{N-2} \c 
{\rm Tr} (\underbrace{\cdots}_{i-1} D \A  \cdots E
) + {\e \over \b} \c {\rm Tr} (\cdots  D \E ) \right \}  \ .    
\label{carcurrent2}
\ee  
Note that the site index $k$ has disappeared from
the expression and thus the current is
independent of $k$, as it should be from the conservation of
the number of cars.
Recalling that $\hat{A}=A$ and $\bt DA=A$, 
the first term becomes
$$
{\e \over \bt} \sum_{i=1}^{N-2}\ \c^{N-1,M-1} {\rm Tr} 
 (\underbrace{\cdots}_{i-1} A \cdots E) \ .
$$
In this double summation, 
each configuration $(\cdots E)$ is counted $[(N-1)-(M-1)-1]$ times
and thus it can be simplified further to
$$
{\e \over \bt} (N-M-1) Y(N-1,M-1)
$$
where 
\be
Y(P,Q) =\c^{P,Q} {\rm Tr} 
(\cdots E) . 
\ee
Also recalling that $\hat{E}=E+e$, 
the second term in Eq.~(\ref{carcurrent2}) can be split
into two pieces
\be
{\e \over \b} Y_D(N,M)+ e{\e \over \b}
\c^{N-1,M}{\rm Tr}(
  \underbrace{\cdots D}_{0 {\rm \ truck} } )
\label{for30.5}
\ee
where
\be
Y_D(P,Q)=\c^{P,Q} {\rm Tr}( 
\cdots DE) . 
\ee
Using $\bt DA= A$ and the cyclic invariance of the trace, 
the second piece in (\ref{for30.5}) can be simplified to
$$
e {\e \over \b}
  {1\over \tilde{\beta}^M}
  C^{M-1}_{N-2} {\rm Tr} (A^{N-M-1}) \ 
$$
where the binomial coefficient $C^{M-1}_{N-2}$ comes from counting 
the number of configurations that satisfy the given specifications.   
Later it turns out that the factor ${\rm Tr}(A^{N-M-1})$
cancels out from all expressions for physical quantities
including $\langle J^{\rm car}_k \rangle$ and 
$\langle J^{\rm truck}_k \rangle$.
Below we thus set ${\rm Tr}(A^{N-M-1})=1$ for convenience.
We also mention that due to the cancellation, all calculations
in this paper can be performed without resorting to
an explicit representation of the matrices.  

To evaluate  the denominator of Eq.~(\ref{CARCURRENT}), 
we proceed with the definition~(\ref{probsum})
\be
Z(N,M)
=\c {\rm Tr}(\cdots {\hat E}) 
+ \c {\rm Tr}(\cdots {\hat D}) + \c {\rm Tr}(\cdots {\hat A}) \ .
\label{totalpartition}
\ee  
In a similar evaluation procedure as above,
the first term on the RHS  yields
\be
\c {\rm Tr}(\cdots {\hat E}) = 
Y(N,M) +  {e \o {\bt ^{M}}}C_{N-1}^{M} \ .
\ee  
Also the sum of the second and the third terms on the RHS becomes
\be
\sum _{i=1}^{N-1} \c \left\{ 
{\rm Tr}(\underbrace{\cdots}_{i-1}  {\hat D} \cdots E) 
+ {\rm Tr}(\underbrace{\cdots}_{i-1}  {\hat A} \cdots E) \right\} \ ,
\ee 
which results in
\be
(N-1)Y(N,M) +d (N-1) Y(N-1,M-1) \ .
\ee  

Combining all calculations given above, one obtains the exact expression
for $\langle J^{\rm car}_k \rangle$ for arbitrary $N$ and $M$:
\begin{eqnarray}
\langle J^{\rm car}_k\rangle &=&
 {1 \o Z(N,M) }  \left\{ \rule{0mm}{8mm}{\e \over \bt}  
  (N-M-1)Y(N-1,M-1) \right.   
  \nonumber \\
 & &  \left. + {\e \over \b} Y_D(N,M) + e {\e \over \b} 
  {1 \o \bt^M} C_{N-2}^{M-1}  
  \right\} \ .
\label{for9}
\end{eqnarray}
For a complete evaluation of Eq.~(\ref{CARCURRENT}), one needs
$\y$ and $\yd$. Exactly same quantities are
calculated in the RSU scheme~\cite{lee}. 
Taking into account the renormalization of $\a$ and $\b$, 
one immediately obtains
\be
Y_D(N,M)= {1\o {\at \bt ^M}} \left [ {\at\o {1- \bt}} C_{N-2}^{M-1} +
{\at +\bt -1 \o {\bt -1}} I(N,M)\right ] 
\label{for11}
\ee
\be
Y(N,M)=Y_D(N,M) +  {1\o {\at \bt ^M}} C_{N-2}^{M} 
  \ \ , \ \ 
I(N,M)= \sum_{k=1}^M \bt^k C_{N-2-k}^{M-k} \ .
\label{for12}
\ee

To study the implications of Eq.~(\ref{for9}), 
we consider the thermodynamic limit $N,M \rightarrow \infty$ 
while  $n=M/ N$ is kept fixed. 
As demonstrated in Ref.~\cite{lee}, 
the thermodynamic limit is governed by $I(N,M)$.
Using the method of steepest descent, it can be
verified that for $n\tilde{\beta}>1$
$$
I(N,M) \rightarrow {\tilde{\b}^{N-1} \over 
  (\tilde{\b}-1)^{N-M-1} } \ ,
$$
which is exponentially larger than all other terms and 
hence dominates the thermodynamic behavior of all quantities.
For $n\tilde{\beta}<1$, on the other hand,
$$
I(N,M) \rightarrow { n\tilde{\b} (1-n)^2
  \over 1-n\tilde{\b} } C^M_N \ ,
$$
which is of the same order as other terms.
A straightforward calculation then shows that 
$\langle v_{\rm car} \rangle ={1 \o n } \langle J^{\rm car}_k \rangle $
is given by
\bea
\langle v_{\rm car}\rangle=  \ \ \left\{  
\begin{array}{ll} \displaystyle      
{\e \over 1-\e n}(1-n) 
  & {\rm if } \ \ \ n\bt \leq 1 \\ 
\displaystyle \rule{0mm}{8mm}
  {\e \over \b -\e}{1-n \over n}
  & {\rm if} \ \ \  n\bt \geq 1  \ .
\end{array} \right. 
\label{v_car_BSU}
\eea

The average current of the single truck [Eq.~(\ref{truckcurrent})]
can be evaluated in a similar manner:
 \begin{eqnarray}
\langle J^{\rm truck}_k \rangle &=& { 1 \over Z(N,M) }
  \left\{ \e \gamma \left[ \y - \yd + 
   {e \o \bt^M} C_{N-2}^M \right]
  \right. \nonumber \\
 & & \left. + {\e \over \b} \left[ Y_D(N,M) + 
   {e \o \bt^{M}} 
  C_{N-2}^{M-1}\right] 
  \right\} \ .  
\end{eqnarray}
In the thermodynamic limit, 
the average velocity 
$\langle v_{\rm truck} \rangle=N \langle J^{\rm truck}_k \rangle $
becomes
\bea
\lefteqn{ \langle v_{\rm truck}\rangle} \nonumber \\
 &= &  \ \ \left\{  
\begin{array}{ll} \displaystyle       
  {\e \over \b}   {
  \left\{ \a (1-n)+ e \at 
     \left[ n+\a (1-n) \right] \right\}  
   (1-n\tilde{\b}) 
  +n(\tilde{\a} + \tilde{\b}-n\tilde{\b})
  \over (1-\eta n) \left[ (1-n)(1-n\tilde{\b}) 
  + n (\tilde{\a} + \tilde{\b}-n\tilde{\b}) \right] }
&  {\rm if } \ \ \ n\bt \leq 1 \\ 
\displaystyle \rule{0mm}{8mm}\frac{\e}{\b - \e} 
  & {\rm if } \ \ \ n\bt \geq 1 \ .
\end{array} \right. 
\eea

Figures 1 and 2 show the behaviors of $\langle v_{\rm car}\rangle$ 
and $\langle v_{\rm truck}\rangle$  as a function of 
$ 1-{1 \o \b} $, the narrowness of the road, 
while the values of $\e , \gamma(={\a \over \b }) $ and $n$ are fixed.
The behavior of $\langle v_{\rm car} \rangle$ changes clearly
at the transition point $1-{1 \o \b_c} =1-n$. 
Below the transition point, $\langle v_{\rm car} \rangle$ is
constant, implying that the interaction with the truck
causes negligible changes on the flow of cars.
Above the transition point, on the other hand, 
$\langle v_{\rm car} \rangle$ begins to drop suddenly
generating a cusp at the transition point. 
Thus a local dynamic ``{\it impurity }'', truck, results in global effects.
Here the decrease of $\langle v_{\rm car} \rangle$ follows
a hyperbolic curve.  
$\langle v_{\rm truck}\rangle$ also changes its behavior
at the transition point. Below the transition point, 
one finds a rather smooth decrease in $\langle v_{\rm truck}\rangle$ 
and above the transition point,
one again observes the hyperbolic decrease.

\subsection{FSU scheme} 

In the FSU scheme, the average current
of cars $\langle J_k^{\rm car} \rangle$ between
the site $k$ and $k+1$ reads
\be
\langle J_k^{\rm car}\rangle= {1\o {Z(N,M)}}
\left\{ \e \c {\rm Tr} (\underbrace{\cdots}_{k-1} \D A  \cdots) 
+ {\e \over \b} \c {\rm Tr} 
  (\underbrace{\cdots}_{k-1} \D E  \cdots) \right \} \ .      
\ee 
In comparison with Eq.~(\ref{CARCURRENT}) for the BSU scheme, 
it should be noticed that the hats now appear at the site $k$ 
instead of $k+1$ since the action of $T_{k+1,k}$ occurs
after the steady state is acted by
$T_{k,k-1} \cdots T_{2,1}T_{1,N}$. This expression
can be derived in a more rigorous way following a similar
procedure as sketched in Appendix A.
Using the quadratic algebra of the matrices, this expression
can be reduced to 
\begin{eqnarray}
 \langle J_k^{\rm car}\rangle &=&{1 \over Z(N,M) }  
  \left\{ \e\left( {1 \over \bt}+d \right) 
  (N-M-1) Y(N-1,M-1) \right. 
  \nonumber \\
 & & \left. + {\e \over \b} \left[ Y_{D}(N,M) 
  + d Y(N-1,M-1) \right] \right\} \ .
\end{eqnarray}
Note that the $k$ dependence has disappeared.
Here it should be understood that $\at,\bt,d,e$ now
have the values given in Eqs.~(\ref{adeFSU},\ref{atbtFSU}). 
The same understanding is also needed for $Y(N,M),Y_D(N,M),Z(N,M)$
whose explicit expressions are given in the preceding subsection.

Similar shifts of the hats occur for the average current 
of the truck as well and one finds
\be
\langle J_k^{\rm truck}\rangle= {1\o {Z(N,M)}}
\left\{ \e \gamma \c {\rm Tr} 
  (\underbrace{\cdots}_{k-1} \A E  \cdots) 
+ {\e \over \b} \c {\rm Tr} 
  (\underbrace{\cdots}_{k-1} \D E  \cdots) \right \} \ ,     
\ee 
which is equal to
\begin{eqnarray}
\langle J_k^{\rm truck}\rangle &=&{1\over  Z(N,M)}  
  \left\{ \rule{0mm}{6mm} \e \gamma \left[ Y(N,M) 
  -Y_{D}(N,M) \right] \right. 
  \nonumber \\
 & & \left. \rule{0mm}{6mm}+{\e \over \b} \left[ Y_{D}(N,M) + 
 d Y(N-1,M-1) \right] \right\} \ .
\end{eqnarray}
Note again that the $k$ dependence has disappeared.

Using the relations 
$\langle v_{\rm car}\rangle={1\over n}\langle J_k^{\rm car}\rangle$ 
and 
$ \langle v_{\rm truck}\rangle=N\langle J_k^{\rm truck}\rangle$, 
and taking the thermodynamic limit, one reaches the following results
\bea
\langle v_{\rm car}\rangle =  \ \ \left\{  \begin{array}{ll}      
\displaystyle 
  {\e (1-n)\o {1-\e (1-n)}} 
  & {\rm if } \ \ \ n\bt \leq 1 \\ 
\displaystyle  \rule{0mm}{9mm}
  {\e \over \b} {(1-n) \o n} & {\rm if } \ \ \ n\bt \geq 1 
\end{array} \right.
\label{v_car_FSU}
\eea
\bea
\langle v_{\rm truck}\rangle=  \ \ \left\{  \begin{array}{ll}      
\displaystyle {\e \over \b} { (\at +dn\bt) (1-n)(1-n\bt)
  +(1+ dn\bt) n(\at+\bt-n\bt)  \over
  \left(1+ dn\bt\right)
  \left[ (1-n)(1-n\bt)+n(\at+\bt-n\bt) \right] } 
 & {\rm if } \ \ \ n\bt\leq 1 \\ 
\displaystyle {\e \over \b} & {\rm if } \ \ \ n\bt \geq 1 \ .
\end{array} \right.
\eea

Figures 3 and 4 show $\langle v_{\rm car}\rangle$
and $\langle v_{\rm truck}\rangle$ 
as a function of the narrowness $1-{1 \o \b} $ 
while $\e,\gamma (={ \a \over \b} )$, and $n$ are fixed. 
One again finds that the average velocities have cusps 
at the transition point. 
Above the critical narrowness, both
$\langle v_{\rm car} \rangle$ and $\langle v_{\rm truck} \rangle$ 
decrease linearly with respect to the narrowness,
which is in contrast to the hyperbolic decreases in the BSU scheme.
Below the critical narrowness, $\langle v_{\rm car} \rangle$ 
is constant, similar to the result in the BSU scheme.
But its value is different from the corresponding one 
in the BSU scheme.

\subsection{Comparison with the RSU results} 

In the thermodynamic limit, $\langle v_{\rm car}\rangle_{\rm RSU}$ 
and $\langle v_{\rm truck}\rangle_{\rm RSU}$ become~\cite{lee}
\bea
\langle v_{\rm car}\rangle_{\rm RSU}=  \ \  \left\{ \begin{array}{ll}      
\displaystyle  1-n & {\rm if } \ \ \ n\b \leq 1 \\ 
\displaystyle \frac{1}{\b}\frac{(1-n)}{n} & {\rm if } \ \ \  n\b \geq 1  
\end{array} \right.
\label{v_car_RSU}
\eea
\bea
\langle v_{\rm truck}\rangle _{\rm RSU}=  \ \ \left\{  \begin{array}{ll}      
\displaystyle {1 \o {\b}}\frac{\a(1-n)(1-n\b) 
  +n(\a + \b-n\b)}{(1-n)(1-n\b) + n (\a + \b-n\b
  )} & {\rm if } \ \ \ n\b \leq 1 \\ 
\displaystyle \rule{0mm}{8mm}\frac{1}{\b} & {\rm if } \ \ \ n\b \geq 1 \ . 
\end{array} \right.  
\eea

We now compare the results. In all updating schemes considered,
a phase transition occurs at a critical value of the narrowness.
In the RSU and BSU schemes, the critical value is $1-n$
while in the FSU scheme, it is ${ (1-\e)(1-n) \o 1-\e (1-n)}$.
Note that the transition point can be significantly lower 
in the FSU scheme when $\eta \approx 1$. 
Below the critical narrowness, $\langle v_{\rm car}\rangle $ 
is independent of $\b$ in all three schemes (with different values 
in each scheme) and above it, $\langle v_{\rm car}\rangle $ decreases 
linearly in the RSU and FSU schemes but 
hyperbolically in the BSU scheme.
$\langle v_{\rm truck}\rangle $, on the other hand, is not
constant even below the critical narrowness in all three schemes  
and varies quite smoothly with respect to the narrowness.
Above the critical narrowness, $\langle v_{\rm truck}\rangle $
decreases linearly in the RSU and FSU schemes 
but hyperbolically in the BSU scheme.
 
In order to illustrate the origin of these differences,
it is instructive to consider the behavior of 
$\langle v_{\rm truck}\rangle $ in the vanishing car density
limit $n \rightarrow 0$. In this limit, 
$\langle v_{\rm truck} \rangle$ in the RSU and FSU schemes approaches 
$\gamma $ and $\e\gamma $, respectively 
while it approaches ${ \e \gamma  \over 1 - \e \gamma }$
in the BSU scheme. This difference can be explained in the following way.
In the absence of any car, the single truck in the RSU scheme
either hops with the probability $\gamma dt$ or stays at
the present site with the probability $(1-\gamma dt)$ during
a time interval $dt$.  
Hence its velocity is equal to its hopping rate $\gamma$.
The situation is similar in the FSU scheme. 
Since the truck moves in the opposite direction of the update,
it either hops one site ahead with the probability $\eta\gamma$
or stays with the probability $(1-\eta\gamma)$.
Therefore, the average velocity reads
$$ \langle v_{\rm truck}\rangle = { {\rm average \ \ distance} \over  
{\rm number \ \ of \ \ time \ \ step} } 
= { 0\times (1- \e \gamma  ) + 1\times (\e\gamma  ) 
\over 1 } =\e \gamma \ .
$$
The situation changes drastically in the BSU scheme. In this case, the
direction of the update and the direction of the truck movement 
are identical and accordingly the truck can be transferred 
by large distances in a single time step. Recalling that 
each hopping occurs with the probabilities $\e \gamma$ 
and the probability of stopping is $(1- \e \gamma)$, 
$\langle v_{\rm truck} \rangle$ can be cast 
into the following form
$$
0\times (1- \e \gamma) + \e \gamma (1- \e \gamma ) 
+ 2  (\e \gamma)^2 (1- \e \gamma  )   
+ 3  (\e \gamma)^3 (1- \e \gamma  )+ \cdots  
$$
which  is simplified to ${ \e \gamma \over 1- \e \gamma }$.

We next examine the fundamental diagrams  
(the relation between the car current and the car density) in
three updating schemes.
The average current of cars is simply related  to the
average velocity via
$ \langle J_{\rm car} \rangle = n \langle v_{\rm car} \rangle $.
Using the thermodynamic behaviors of $\langle v_{\rm car} \rangle$
[Eqs.~(\ref{v_car_BSU},\ref{v_car_FSU},\ref{v_car_RSU})], 
one can determine the thermodynamic behaviors of
the current as a function of the density. 
Figure 5 shows these
behaviors for some constant values of $\beta$ and $\eta$ in different 
updating schemes. In all three types of update,  
the single truck does not affect  the current-density 
diagram  before the transition point 
whereas  it affects the system after the transition point in a 
nontrivial manner: 
linear decrease of the current with increasing density.

\section{Density Profile}

\subsection{Density profile in the RSU scheme} 

Here we first summarize the results in Ref.~\cite{lee}.
In the RSU scheme, it can be assumed without loss of generality
that the single truck is at a particular site, 
for example at the site $N$ of the chain
due to the cyclic invariance of the problem.
Thus the probability to find
a car at the distance $x$ from the truck can be written via MPS as follows
\be
\langle n(x)\rangle_{\rm RSU} =\frac{\c {\rm Tr}
 (\cdots D \overbrace {\cdots}^x E)}{\c {\rm Tr}(\cdots \cdots E)} \ .
\label{for10}
\ee
Using the matrix algebra (\ref{for1}), one finds 
\begin{eqnarray}
\lefteqn{\langle n(x)\rangle_{\rm RSU} =
  \left [Y_{\rm RSU}(N,M) \a \b ^M\right]^{-1}
 \left\{C_{N-3}^{M-1} - \frac{\a}{\b -1}C_{N-3}^{M-2}
  + \frac{\a + \b -1 }{\b -1}
 \right. } 
 \nonumber \\
 & & \left. \times \rule{0mm}{8mm} 
  \left [ I_{\rm RSU}(N-1,M-1) 
  +\b ^{x -1}(\b -1) I_{\rm RSU}(N-x,M-x) 
  \theta (M \geq x) \right ] \right\}
\label{for13}
\end{eqnarray}
where $Y_{\rm RSU}(N,M)$ and  $I_{\rm RSU}(N,M)$ can be
obtained from $Y(N,M)$ and $I(N,M)$ [Eqs.~(\ref{for11},\ref{for12})]
by replacing $\tilde{\a},\tilde{\b}$ with $\a,\b$, respectively and
the factor $\theta(y\ge x)$ is 1 if
$y\ge x$ and 0 otherwise.
In the high density  phase $n\b\geq 1$, 
the single truck affects the system globally. 
The density profile is 
\bea
\langle n(x)\rangle_{\rm RSU} = \ \ \left\{  \begin{array}{cl} 
\displaystyle 1  & \mbox{for } {x \o N} 
  \le l_{\rm RSU} \equiv  {{n\b -1}\o {\b -1}}  \\ 
\displaystyle \rule{0mm}{8mm}{1\o {\b}}  & {\rm otherwise} \ .
\end{array} \right.
\label{nhighRSU}
\eea
Note that the system consists of two regions, 
a traffic jam region in front of the truck  
and a free flow region behind it. 
In the low density phase $n\b\le 1$, on the other hand,
the presence of the truck 
has only  local effects. In the thermodynamic limit,
the car density becomes 
\be
\langle n(x) \rangle_{\rm RSU} = 
n \left \{ 1+ \frac{(\a + \b -1)(1-n)}{1-n+\a n}(n\b)^x 
  \right \} \ 
\label{nlowRSU}
\ee
which shows that the disturbance by the truck decays
exponentially with a characteristic length scale
\be
\xi_{\rm RSU} =|\ln(n\b)|^{-1} \ .
\ee

\subsection{Density profile in the ordered sequential updating schemes}

As stated in Section 2, the ordered sequential updates
break the cyclic invariance.
In what follows we show that 
due to the absence of the cyclic invariance,
the probability to find a car at $x$ sites in front
of the truck varies depending on the truck location.
Two updating schemes, BSU and FSU, will be considered simultaneously
since same expressions apply to both schemes. 

We first consider the case where the truck is at the site $N$.
The conditional probability to find a car at $x$ sites
in front of the truck reads 
\begin{equation}
{\rm Prob}(N-1-x={\rm car}| N={\rm truck})
 ={{\c {\rm Tr}(\cdots D\overbrace{\cdots}^x \E)}\o 
    {\c {\rm Tr}(\cdots \E)}} \ .
\label{condprob1}
\end{equation}
Here it is helpful to define a quantity $K(x,N,M)$ by
$$
K(x,N,M)={\c {\rm Tr}
 (\cdots D \overbrace {\cdots}^x E) \over \c {\rm Tr}(\cdots \cdots E)} \ 
$$
which is exactly the same as Eq.~(\ref{for10}). All properties
of $K(x,N,M)$ can be thus obtained from Eq.~(\ref{for13}) 
by replacing $\a$ and $\b$ with $\tilde{\a}$ and $\tilde{\b}$,
respectively. In terms of $K(x,N,M)$, 
the conditional 
 
probability~(\ref{condprob1}) becomes

\be
 {{Y(N,M)K(x,N,M)  
  +{e \o \bt^M} C_{N-2}^{M-1} }
 \o {Y(N,M) 
 + {e \o \bt^M} 
  C_{N-1}^{M}}} \ .
\label{for17}
\ee

The thermodynamic limit can be investigated in a simple way.
In the high density phase $n\tilde{\beta} \ge 1$, 
the second terms both in the numerator and the denominator
in Eq.~(\ref{for17}) are negligible compared to
the first terms and Eq.~(\ref{for17}) reduces to $K(x,N,M)$.
Thus one finds
\bea
{\rm Prob}(N-x-1={\rm car}|N={\rm truck})=\ \ \left\{ \begin{array}{cl}
\displaystyle 1  & \mbox{for }  {x\o N} \leq l\equiv 
  {n\tilde{\beta} -1 \over \tilde{\beta}-1} \\ 
\displaystyle {1 \o {\bt}} &  {\rm otherwise} \ 
\end{array} \right. 
\eea
which is essentially identical to the result (\ref{nhighRSU})
in the RSU scheme except for the replacement of $\b$ by $\tilde{\b}$.
In the low density phase $n \bt \leq 1$, on the other hand,
the second terms are comparable with the first terms and one obtains 
\be
{\rm Prob}(N-x-1={\rm car}|N={\rm truck})
=n\left \{ 
 1+ {(\tilde{\a} + \tilde{\b} -1)(1-n) \over
  (1-n+\tilde{\a} n)+e(1-n\tilde{\b}) \tilde{\alpha}}
  (n\tilde{\b})^x 
\right \} \ 
\ee
which shows the exponential decay of the disturbance
with a length scale $\xi=\left|\ln\left(n \bt \right)\right|^{-1}$. 
The coefficient of the exponential decaying term 
is different from the result in the RSU scheme. 

Secondly we consider the case where the truck is 
at the site $N-k$ $(1\le k \le N-x-2)$.
The conditional probability becomes
\be
{\rm Prob}(N-x-k-1={\rm car}|N-k={\rm truck})
={ \c {\rm Tr}(\cdots D \overbrace {\cdots}^x E 
\overbrace {\cdots}^{k-1} {\hat X_N}) \o
{\c {\rm Tr}(\cdots E \underbrace {\cdots}_{k-1} {\hat X_N})}} \ .
\label{for14}
\ee
The evaluation of Eq.~(\ref{for14}) using the matrix algebra leads to
\be
{Y(N,M)K(x,N,M)  +d Y(N-1,M-1)K(x,N-1,M-1)
\over {Y(N,M) +d Y(N-1,M-1)}}   
\label{for18} \ .
\ee
Note that the conditional probability is independent of $k$.

In the thermodynamic limit, (\ref{for18}) can be greatly 
simplified to $K(x,N,M)$ since $K(x,N,M)$ and $K(x,N-1,M-1)$ 
in the numerator become identical in this limit 
[see Eqs.~(\ref{nhighRSU},\ref{nlowRSU})].
Thus the conditional probability 
${\rm Prob}(N-x-k-1={\rm car}|N-k={\rm truck})$
is the same as the corresponding 
results (\ref{nhighRSU},\ref{nlowRSU})
in the high and the low density phases
except for the renormalization of $\a$ and $\b$
to $\at$ and $\bt$.

In the third case, the truck is located 
at the site $x+1$ and the car at the site $N$.
The conditional probability becomes
\be
{\rm Prob}(N={\rm car}| x+1={\rm truck})= 
{{\c {\rm Tr}(\overbrace{\cdots}^x E \cdots \D)}\o 
 {\c {\rm Tr}(\underbrace{\cdots}_x E 
\cdots {\hat X}_N)}} \ ,
\label{condprob3}
\ee
which is found to be
\be
 {{Y(N,M)K(x,N,M)  +d Y(N-1,M-1)
 }\over {Y(N,M) +d Y(N-1,M-1) }} \ .
\label{for19}
\ee
Note that the second terms both in the numerator and the denominator
are comparable to the first terms in the thermodynamic limit. 
As a result, the conditional probability becomes noticeably 
different from $K(x,N,M)$. One finds in the high density phase 
\bea
{\rm Prob}(N={\rm car}|x+1={\rm truck})= \ \ \left\{  \begin{array}{ll}
\displaystyle 1 \ \ &\mbox{for }{x \o N} \leq l = {n\bt -1 \o \bt -1 }\\ 
\displaystyle {1 \over \bt}{1+d\bt \over 1+d}
& {\rm otherwise}
\end{array} \right.
\label{condprob3high}
\eea
and in the low density phase
\be
{\rm Prob}(N={\rm car}|x+1={\rm truck})=
{n(1+d\bt) \over 1+dn\bt } 
\left \{ 1+ {(\at+\bt-1)(1-n) \over
  (1+d\bt)(1-n+\at n)}
\left(n\bt\right)^x \right \} \ .
\ee
In contrast to the previous cases, the renormalization of $\a$ and $\b$
are not sufficient to account for deviations from the RSU results
even in the high density phase.

Lastly we consider ${\rm Prob}(N-l={\rm car}|x-l+1={\rm truck})$ 
$(1\le l\le x)$, which takes the following form
\be
{Y(N,M)K(x,N,M) +d Y(N-1,M-1)K(x-1,N-1,M-1)
 \over Y(N,M) +d Y(N-1,M-1) } \ .
\label{for20}
\ee
Note that it is independent of $l$.
The thermodynamic limit can be examined in a  similar way.
In the high density phase, $K(x,N,M)$ and $K(x-1,N-1,M-1)$ are
essentially identical and thus Eq.~(\ref{for20}) reduces
to Eq.~(\ref{nhighRSU}) for
the RSU scheme except for the trivial
replacement of $\a$ and $\b$ with $\at$ and $\bt$.
In the low density phase, however, the difference between
$K(x,N,M)$ and $K(x-1,N-1,M-1)$ is not negligible and
careful treatment is required to take care of 
the difference. This way, one obtains
\be
{\rm Prob}(N-l={\rm car}|x-l+1={\rm truck})=
n \left \{ 1+ {1+d \o 1+d n \bt  } 
{ (\at +\bt -1) (1-n) \o 1-n+\at n }
\left(n\bt\right)^x \right \} \ .
\ee

We next discuss the origin of the absence of the cyclic invariance,
which results in the four different cases.
As stated above, the system loses the cyclic invariance
due to the choice of a particular site, the site $N$, 
as a starting point of the update.
 
Thus by choosing the starting point in an even way,
the cyclic invariance can be restored.
One can, for example,
define a cyclically invariant 
average $\langle\langle \cdots \rangle\rangle$ of an operator $\hat{\cal O}$ 
in the following way\footnote{This problem due to
the absence of the cyclic invariance does not affect
the calculation of the currents since the currents should
be independent of the sites where their values are evaluated
due to the conservation of the particle number.}: 
\be
\langle\langle{\hat O}\rangle\rangle 
= {1\o N} \sum_{k=1}^N \langle s| {\hat O}|P_{s,k}\rangle \ 
\label{for25}          
\ee
where $\langle s|=\sum_{\{\tau\}}\langle \{\tau\}|$ and
$|P_{s,k} \rangle _k$ is the stationary state of 
the transfer matrix
$$
T_{\leftarrow,k}\equiv T_{k,k+1}T_{k+1,k+2}
\cdots T_{N,1}T_{1,2} \cdots T_{k-1,k} \ 
$$
in the BSU scheme or
$$
T_{\rightarrow,k}\equiv T_{k,k-1}T_{k-1,k-2}
\cdots T_{1,N}T_{N,N-1} \cdots T_{k+1,k} \ 
$$
in the FSU scheme with the site $k$ as a starting point of the update.
Note that the site $N$ now loses its special meaning and
the cyclic invariance becomes evident in Eq.~(\ref{for25}).
It is also worth mentioning that the definition~(\ref{for25})
is equivalent to taking an expectation value
at each sub-step of an update, that is,  
after the action of $T_{l,l+1}$ or $T_{l+1,l}$ instead of
a single whole step of the update $T_{\leftarrow}$ or $T_{\rightarrow}$,
and taking the average over these expectation values.

Using the definition~(\ref{for25}), we calculate the density 
$\langle\langle n(x) \rangle\rangle$ 
$=$ $\langle\langle \delta_{\tau_{N-l},2} \delta_{\tau_{N-l-x-1},1} 
 \rangle\rangle$
$/ \langle\langle \delta_{\tau_{N-l},2} \rangle\rangle$ 
 $(0\le l \le N-1)$,
where the sites $0,-1,-2,\cdots$ should be identified with
the sites $N,N-1,N-2,\cdots$, respectively. 
The density $\langle\langle n(x) \rangle\rangle$ is, by definition, 
independent of the truck location $N-l$ 
and thus one may choose $l=0$.
The denominator $\langle\langle \delta_{\tau_{N},2} \rangle\rangle$ 
is ${1 \o N} $
due to the cyclic invariance and the numerator 
$\langle\langle \delta_{\tau_{N},2} \delta_{\tau_{N-x-1},1} 
\rangle\rangle$ becomes
\begin{eqnarray}
{1 \over NZ(N,M)}\left\{ \c {\rm Tr}(\cdots D \underbrace{\cdots}_{x} \hat{E})
 + \sum_{k=0}^{N-x-3}\c {\rm Tr}(\underbrace{\cdots}_{k}\hat{X}_{k+1}
  \cdots D \underbrace{\cdots}_{x}E)  \right. \nonumber \\
 \left. +\c {\rm Tr}(\cdots \hat{D}\underbrace{\cdots}_{x}E)
 + \sum_{l=0}^{x-1} \c {\rm Tr}( \cdots D \underbrace{\cdots}_{l}
  \hat{X}_{N-x+l} \underbrace{\cdots}_{x-l-1}E) \right\} \ .
\label{cyclicdensity}
\end{eqnarray}
The cyclically invariant density then becomes
\begin{eqnarray}
\langle\langle n(x)\rangle\rangle &=&{1 \over Z(N,M)}\left[
  NY(N,M)K(x,N,M)+{e \over \bt^M}C^{M-1}_{N-2}+dY(N-1,M-1) \right.
  \nonumber \\
& &  \rule{0mm}{6mm}+d(N-x-2)Y(N-1,M-1)K(x,N-1,M-1)
  \\
& & \left. \rule{0mm}{6mm}+dxY(N-1,M-1)K(x-1,N-1,M-1) \right]
  \nonumber \ 
\end{eqnarray}
which, in the thermodynamic limit, reduces to
the results identical to the thermodynamic behaviors
of the second conditional probability
${\rm Prob}(N-x-k-1={\rm car}|N-k={\rm truck})$ $(1 \leq k \leq N-x-2)$.

\section{Density-density correlation function}

Here we calculate the two-point equal time correlation function
of the car density using the cyclically invariant
average~(\ref{for25}). Both BSU and FSU schemes are 
considered simultaneously. In terms of the MPS, the density-density
correlation function becomes ($x_1<x_2$ is assumed)
\begin{eqnarray}
& & \langle\langle n(x_1)n(x_2) \rangle\rangle ={1 \over Z(N,M) } 
  \left[ \c {\rm Tr}(\cdots D \overbrace{ \cdots D
  \underbrace{\cdots}_{x_1} }^{x_2}\hat{E} )
  \right. \nonumber \\
& & \ \ \displaystyle 
  +\sum_{k=0}^{N-x_2-3}\c {\rm Tr}(\cdots D \overbrace{ \cdots D
  \underbrace{\cdots}_{x_1} }^{x_2}E \underbrace{\cdots}_k \hat{X}_N ) 
  +\c {\rm Tr}(\cdots D \underbrace{ \cdots}_{x_1} E
  \underbrace{\cdots}_{N-x_2-2} \hat{D} )
  \\
& & \ \ \displaystyle 
 +\sum_{k=0}^{x_2-x_1-2}\c {\rm Tr}(\cdots D \underbrace{ \cdots}_{x_1} E
  \underbrace{\cdots}_{N-x_2-2} D \underbrace{\cdots}_k \hat{X}_N )
  +\c {\rm Tr}( \underbrace{ \cdots}_{x_1} E
  \underbrace{\cdots}_{N-x_2-2} D \cdots \hat{D} )
  \nonumber \\
& & \ \ \displaystyle \left.
  +\sum_{k=0}^{x_1-1}\c {\rm Tr}( \cdots E \underbrace{ 
  \overbrace{\cdots}^{N-x_2-2} D \cdots }_{N-x_1-2} D 
  \underbrace{\cdots}_k \hat{X}_N )
\right] \ .
 \nonumber 
\end{eqnarray}
After some algebra, it can be verified that
\be
\langle\langle n(x_1)n(x_2) \rangle\rangle =
\langle\langle n(x_2)\rangle\rangle -f(x_1)
\ee
where $f(x_1)$ is given by
\begin{eqnarray}
f(x_1) &=& {1 \over Z(N,M)} \left\{ {e \over \bt^M} C^{M-1}_{N-3}
  +d Y(N-1,M-1)\left[ 1 -K(x_1,N-1,M-1)\right] \right. 
  \nonumber \\
  & &  +{1\over \bt}NY(N-1,M-1)\left[1-K(x_1,N-1,M-1)\right] 
    \label{f_x} \\
  & &  + {d\over \bt}(N-x_1-3)Y(N-2,M-2)\left[1-K(x_1,N-2,M-2)\right]
   \nonumber \\
  & & \left. +{d\over \bt} x_1 Y(N-2,M-2)\left[1-K(x_1-1,N-2,M-2)\right] 
   \right\}
   \nonumber \ .
\end{eqnarray}

The connected part of the two-point correlation function,
$\langle\langle n(x_1)n(x_2) \rangle\rangle_C$
$=$ $\langle\langle n(x_1)n(x_2) \rangle\rangle
-\langle\langle n(x_1) \rangle\rangle 
 \langle\langle n(x_2) \rangle\rangle$,
can be used to estimate the degree of correlation.
In the thermodynamic limit, one finds
\be
\langle\langle n(x_1)n(x_2) \rangle\rangle_C=
\left[\langle\langle n(x_2) \rangle\rangle-\kappa \right]
  \left[1-\langle\langle n(x_1) \rangle\rangle \right]
\ee
where $\kappa ={\rm min}(n,{1 \over \bt})$. 
In the high density phase, the connected part
has non vanishing value only when both
$x_1$ and $x_2$ are within the region
$[Nl-\sqrt{N}\Delta,Nl+\sqrt{N}\Delta]$
where $\Delta=\sqrt{2\bt(1-n)}/(\bt-1)$. 
In the low density phase, the connected part
has non vanishing value only when $x_1<x_2 \sim \xi$.
Thus one concludes that the correlation develops
only in the region where the density variation occurs,
which is identical to the conclusion in the RSU scheme \cite{lee}.

\section{Two Trucks and Bound State }

In this section we examine the system with two trucks and $M$ cars. 
We determine the probability $\Omega(R)$ that 
the distance between the trucks is  $R  \  [0\leq R\leq (N-1)/  2]$. 
In the RSU scheme, the cyclic invariance of the MPS allows one to set one
of the trucks at the site $N$ and one obtains
$$
\Omega_{\rm RSU} (R) \sim \c {\rm Tr}(\cdots E\underbrace {\cdots}_RE)
$$
where the sum runs over all configurations with $M$ cars 
and two trucks. Its thermodynamic limit is examined in Ref.~\cite{lee}.
In the low density phase $n\b \leq 1$, it becomes 
\be
\Omega_{\rm RSU} (R) \sim  
  1+ { n(1-n)(\a +\b -1)(\a -1) \o (1-n+\a n)^2 }(n\b)^R \ ,
\ee
which is maximal at $R=0$ and decays exponentially 
with the same length scale
$\xi_{\rm RSU} =|\ln(n\b)|^{-1}$ as in the density profile. 
In the high density phase $n\b \geq 1$,
the probability decreases linearly with $R$  for $0\leq R\leq Nr_{\rm RSU}$, 
where $r_{\rm RSU}={\rm min}(l_{\rm RSU},1-l_{\rm RSU})$
$(r_{\rm RSU} < {1 \over 2} ) $,
and  remains constant
for $Nr_{\rm RSU} \leq R \leq (N-1)/2$. 
The relative ratios are:
\be
{\rm for} \ \ \  r_{\rm RSU}=l_{\rm RSU}, \ \ \ 
{{\Omega_{\rm RSU} (Nr_{\rm RSU})}\o {\Omega_{\rm RSU} (0)}} 
= 1- {{(\a -1)(\b -1)}\o
{\a \b }}
\ee
and
\be
{\rm for}  \ \ \ r_{\rm RSU}=1-l_{\rm RSU}, \ \ \ 
{{\Omega_{\rm RSU} (Nr_{\rm RSU})}\o {\Omega_{\rm RSU} (0)}} 
= 1- {{(\a -1)(\b -1)}\o
{\a \b }}{r_{\rm RSU} \o {1-r_{\rm RSU} }} \ .
\ee
One can interpret this as the formation of a weak bound state 
between the two trucks. 

Now we consider $\Omega (R)$ in the BSU and FSU schemes.
In terms of the matrix products, $\Omega(R)$ in both schemes
can be expressed (up to a proper normalization constant) as
\begin{eqnarray}
& \displaystyle \sum_{k=0}^{N-R-3}\c {\rm Tr}
(\cdots E\underbrace{\cdots}_{R}E\underbrace{\cdots}_{k} \hat{X}_N)
+ \c {\rm Tr}
(\underbrace{\cdots}_{N-R-2}E\underbrace{\cdots}_{R}\hat {E}) 
& \nonumber \\
& \displaystyle + \sum_{k=0}^{R-1}\c {\rm Tr}
(\cdots E\underbrace{\cdots}_{N-R-2}E\underbrace{\cdots}_{k}\hat{X}_N)      
+ \c {\rm Tr}
(\underbrace{\cdots}_{R}E\underbrace{\cdots}_{N-R-2}\hat {E}) & \ .      
\label{B.1}
\end{eqnarray}
It is useful to introduce a quantity $W(N,M,R)$ which is defined by
\be
W(N,M,R)= \c {\rm Tr}(\cdots E\underbrace{\cdots}_R E) \ .
\label{B.3}
\ee 
Then using the relation $ \A =A ,\D =D+d, \E = E+e$ 
and the cyclic invariance of the trace, 
the expression~(\ref{B.1}) can be written in terms of $W(N,M,R)$
as follows:
\begin{eqnarray}
NW(N,M,R)+
d(N-R-2)W(N-1,M-1,R)
\nonumber \\
+dRW(N-1,M-1,R-1)
+2e Y (N-1,M) \ .
\label{B.4}
\end{eqnarray}

The thermodynamic limit can be investigated in a simple way
using the fact that in the RSU scheme, $W(N,M,R)$ is 
identical to $\Omega_{\rm RSU}(R)$ up to a normalization factor.
Then through the renormalization of $\a$ and $\b$, 
its $R$ dependence in the BSU and FSU schemes can be obtained.
Also the last term in the expression~(\ref{B.4}) 
is negligible compared to the first three terms. 
In the high density phase, one then finds that
$W(N,M,R)$, $W(N-1,M-1,R)$, and $W(N-1,M-1,R-1)$ are all 
proportional to each other and thus $\Omega (R)$ 
can be obtained from $\Omega_{\rm RSU}(R)$ through
the ``{\it renormalization}'' of $\a$ and $\b$.
In the low density phase,
the $R$ dependence of $W(N,M,R)$ appears only
for $R\sim \xi$. It is then sufficient to examine
the case $R\sim \xi \ll N$, where the first and the second
terms are dominant and give the same $R$ dependence. 
Hence  $\Omega (R)$ can be again obtained from $\Omega_{\rm RSU}(R)$ by
replacing $\a$ and $\b$ by their ``{\it renormalized}''
values.

\section{Concluding remarks}

We have investigated the characteristics of an exactly solvable two-way
traffic model with the ordered sequential updates (OSU) 
and observed both qualitative and quantitative differences 
in the properties of model from the results obtained with 
the random sequential update \cite{lee}. Our approach is based
on the so-called matrix product formalism which allows analytic solutions. 
In the OSU schemes, the choice of a particular site as a starting
point of the update breaks the translational invariance of 
the steady state measure, which is also evident in the form 
of the MPS.  Thus an averaging over the different choices
of the update starting point is necessary to restore 
the cyclic invariance to the system [see Eq.~(\ref{for25}) and 
the following discussion]. Performing the cyclically invariant averaging,
some characteristics in the thermodynamic limit, such as 
density profile of cars, density-density correlation function,
and truck-truck distance distribution $\Omega (R)$, are obtained
and it is found that the difference in the updating schemes can be
taken into account simply by 
the proper renormalization of the parameters
$\a$ and $\b$. However this is not the case with average velocities. 
Changing the update scheme affects velocities 
in a more complicated manner and the renormalization of
the parameters is not sufficient to account for
different behaviors of $\langle v_{\rm car } \rangle $ and 
$\langle v_{\rm truck } \rangle $ in different updating schemes.
Especially the dependence of $\langle v_{\rm car} \rangle$
and $\langle v_{\rm truck} \rangle$ on the road narrowness
$1-{1\over \b}$ can vary qualitatively depending 
on the updating schemes. 
Behaviors of  $\langle v_{\rm car} \rangle$
and $\langle v_{\rm truck} \rangle$ in the FSU scheme
(that is, when the update direction is parallel with
the movement direction of the majority of vehicles)
have more  resemblance to the RSU results than those in
the BSU scheme. In the FSU scheme, however, one observes a shift 
in the value of the critical narrowness
from $1-n$ to ${(1-\eta)(1-n) \over 1-\eta (1-n)} $, which
can be considerable if $\e\approx 1$.

\vspace{1cm} 
\noindent {\bf Acknowledgments} \\
 
M.E.F. would like to thank V. Karimipour for fruitful comments 
and D. Kim, R. Asgari for useful helps. H.-W.L. thanks D. Kim for
bringing his attention to this problem. 
H.-W.L. was supported by the Korea Science and Engineering Foundation
through the fellowship program and the SRC program at SNU-CTP.

\appendix

\section{Derivation of the expression (\ref{CARCURRENT})}

Here we derive the expression (\ref{CARCURRENT})
for the average current of cars in the BSU scheme.
Similar procedures can be used to obtain average currents
of the truck and also the average currents in the FSU scheme.
The starting point is the continuity equation, 
which in discrete time dynamics takes the form 
\be
\langle n_k ^{\rm car}\rangle_{j+1} -  \langle n_k ^{\rm car}\rangle_{j} = 
  \langle J_{k-1}^{\rm car}\rangle_{j} -
  \langle J_{k}^{\rm car}\rangle_{j}
\label{A.1}
\ee
where $\langle \cdots \rangle_j$ represents
the average at the time step $j$.
In terms of the initial state of the system $|P,0\rangle$, 
the LHS becomes
\be
\langle s|n_k^{\rm car} T_{\leftarrow}T^j_{\leftarrow}|P,0\rangle  - 
\langle s|n_k^{\rm car} T_{\leftarrow}^j|P,0\rangle 
\label{A.2}
\ee   
where the bra vector $\langle s|$ is defined by 
\be
\langle s|=\sum_{\{\t\}}\langle \t_1|\ot \langle\t_2| \ot \cdots
 \ot \langle \t_N| \ .
\label{A.3}
\ee  
The conservation of the probability ensures that 
$\langle s|T_{\leftarrow}= \langle s|$ \cite{schutzbook}, which  
then allows one to rewrite (\ref{A.2}) as 
$\langle s|[n_k^{\rm car},T_{\leftarrow} ] T^j_{\leftarrow} |P,0\rangle $. 
Next we evaluate the commutator $[n_k^{\rm car},T_\leftarrow]$.
Using the relation
$$
n_k^{\rm car}=
 \underbrace{ {\bf 1}\ot {\bf 1} \ot \cdots \ot {\bf 1} }_{k-1}\ot 
 |1\rangle\langle 1| \ot 
 \underbrace{ {\bf 1} \ot {\bf 1} \ot \cdots \ot {\bf 1} }_{N-k} \ ,
$$
one finds
\begin{eqnarray}
& \displaystyle  T_{N,1} \cdots  T_{k-2,k-1} \left(\e E_{k-1}^{0,1}
E_{k}^{1,0} +{\e \over \b} E_{k-1}^{2,1} E_{k}^{1,2} \right)
 T_{k,k+1} \cdots T_{N-1,N} & 
 \nonumber \\
 & \displaystyle  - T_{N,1} \cdots T_{k-1,k}
\left( \e E_{k}^{0,1} E_{k+1}^{1,0} 
+ {\e \over \b} E_{k}^{2,1} E_{k+1}^{1,2} \right) T_{k+1,k+2}
\cdots T_{N-1,N} &
\label{A.4}
\end{eqnarray} 
where
$$E_k^{i,j}=
 \underbrace{ {\bf 1}\ot {\bf 1} \ot \cdots \ot {\bf 1} }_{k-1}\ot 
 |i\rangle\langle j| \ot 
 \underbrace{ {\bf 1} \ot {\bf 1} \ot \cdots \ot {\bf 1} }_{N-k} \ .
$$
Comparison with the RHS of Eq.~(\ref{A.1}) (note that the two
currents have different subscripts $k-1$ and $k$) shows that
each term in (\ref{A.4}) should lead to 
the expression for $\langle J^{\rm car}_{k-1} \rangle_j$
and $\langle J^{\rm car}_{k} \rangle_j$, respectively.
Then by taking the limit $j\rightarrow \infty$ and
using the expression (\ref{for4}) for
the steady state 
$|P_s\rangle=\lim_{j\rightarrow \infty} T^j_\leftarrow |P,0\rangle$, 
one finds 
\begin{eqnarray}
\langle J_k^{\rm car}\rangle  &=& 
{1\o {Z(N,M)}} \langle s| 
\left( \e E_{k}^{0,1} E_{k+1}^{1,0} 
+ {\e \over \b} E_{k}^{2,1} E_{k+1}^{1,2}\right)  \nonumber \\
& & \times {\rm Tr}\left (
\underbrace{|U\rangle \ot |U \rangle \ot \cdots |U\rangle}_k \ot 
|\hat{U}\rangle \ot |U\rangle \ot \cdots \ot |U\rangle \right)  \ 
\label{A.5}
\end{eqnarray}
where the effects of $T_{l,l+1}$ on $|P_s\rangle$ and
$\langle s|$ have been taken into account.
Finally we use
$$
|i\rangle \langle j| \left( A_0 |0\rangle +
  A_1 |1\rangle + A_2 |2\rangle \right)=
 A_j |i\rangle \ ,
$$
which yields the expression (\ref{CARCURRENT}).



\begin{figure}
\caption{Average velocity of cars
  in the BSU and RSU schemes for $n=0.3$
and different values of $\eta$ }
\end{figure}

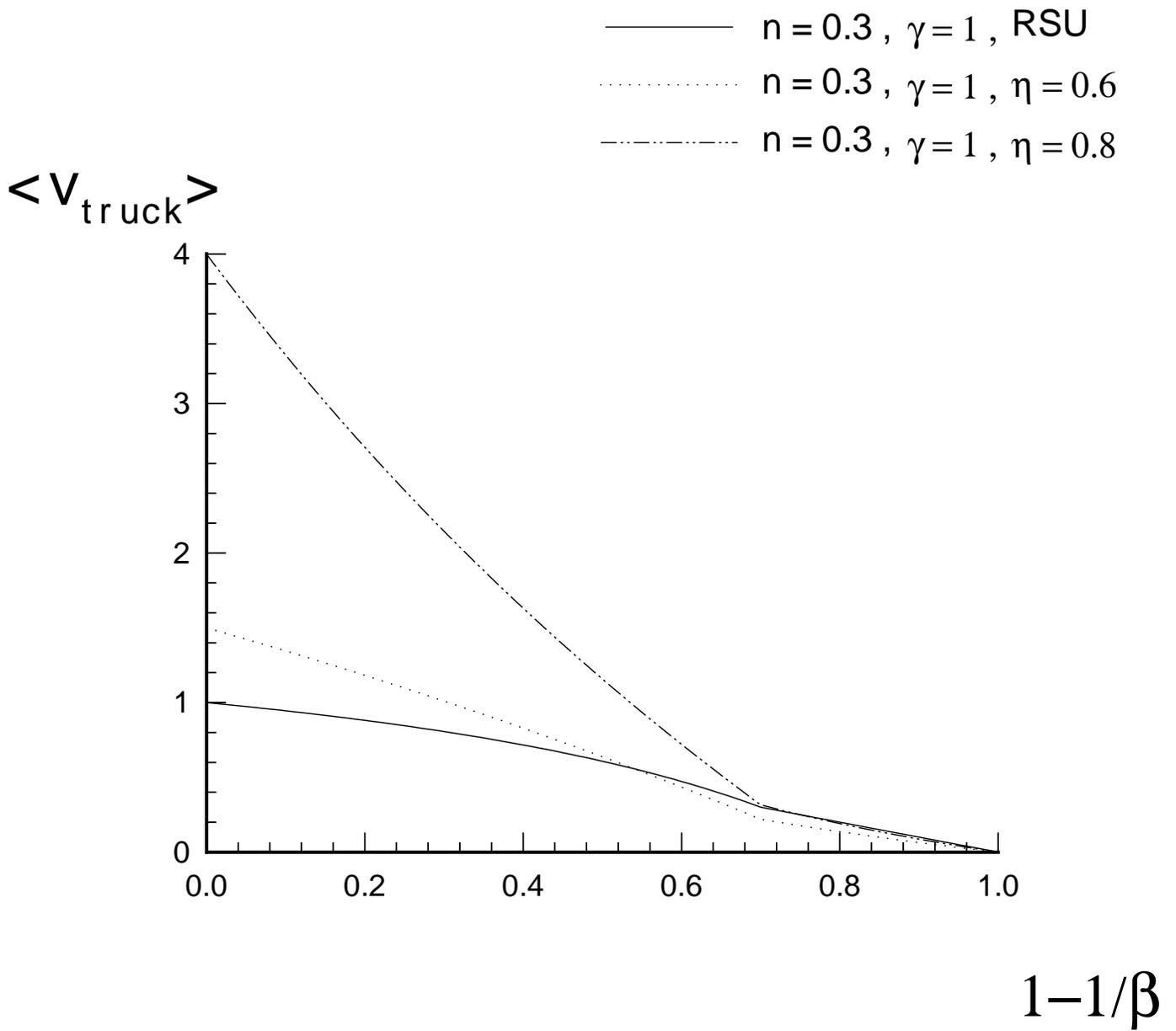
\begin{figure}
\caption{Average velocity of the truck 
 in the BSU and RSU schemes for
$n=0.3$ and different values of $\eta$ }
\end{figure}

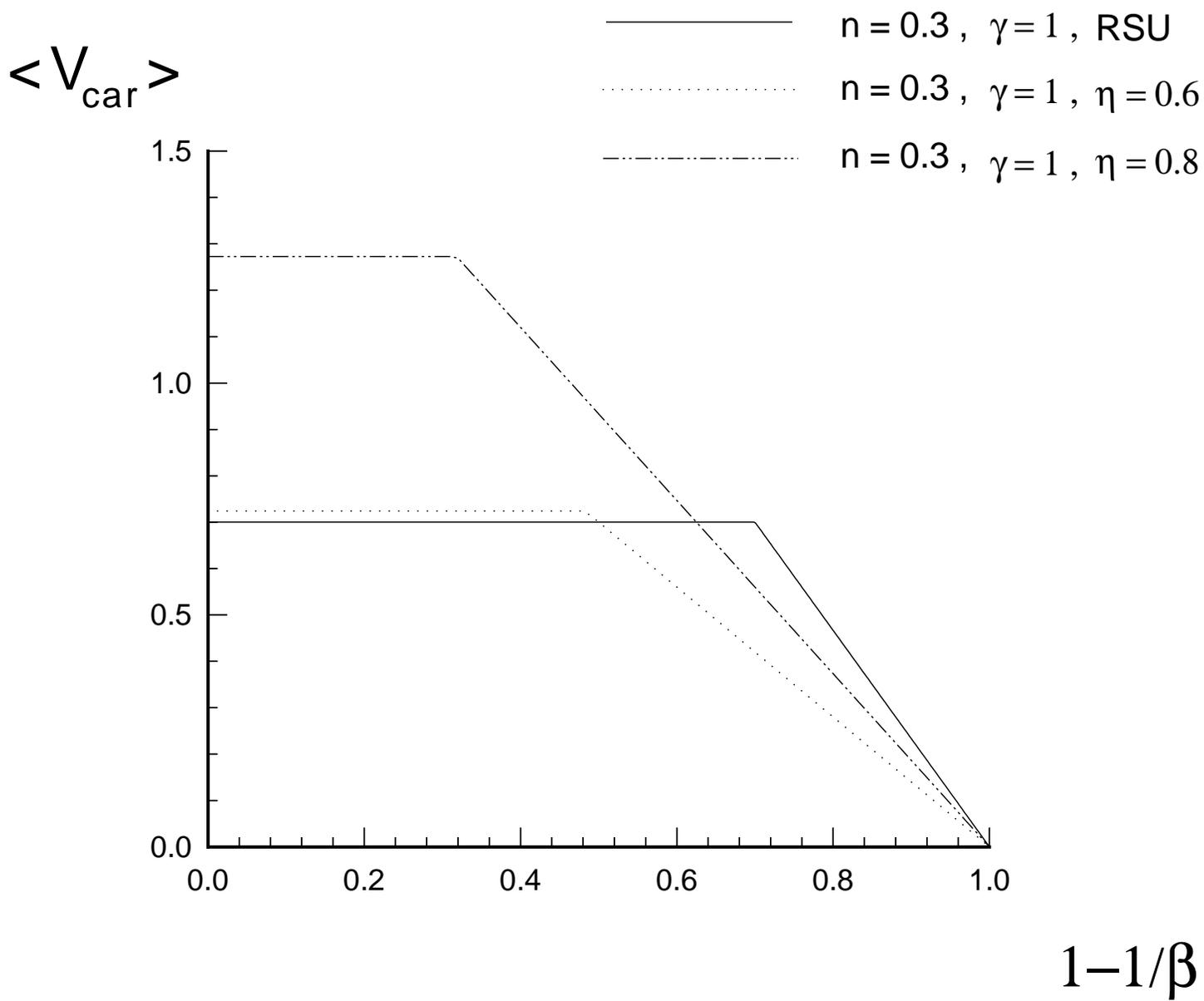
\begin{figure}
\caption{Average velocity of cars in the FSU and RSU schemes for $n=0.3$
and different values of $\eta $ }
\end{figure}
 
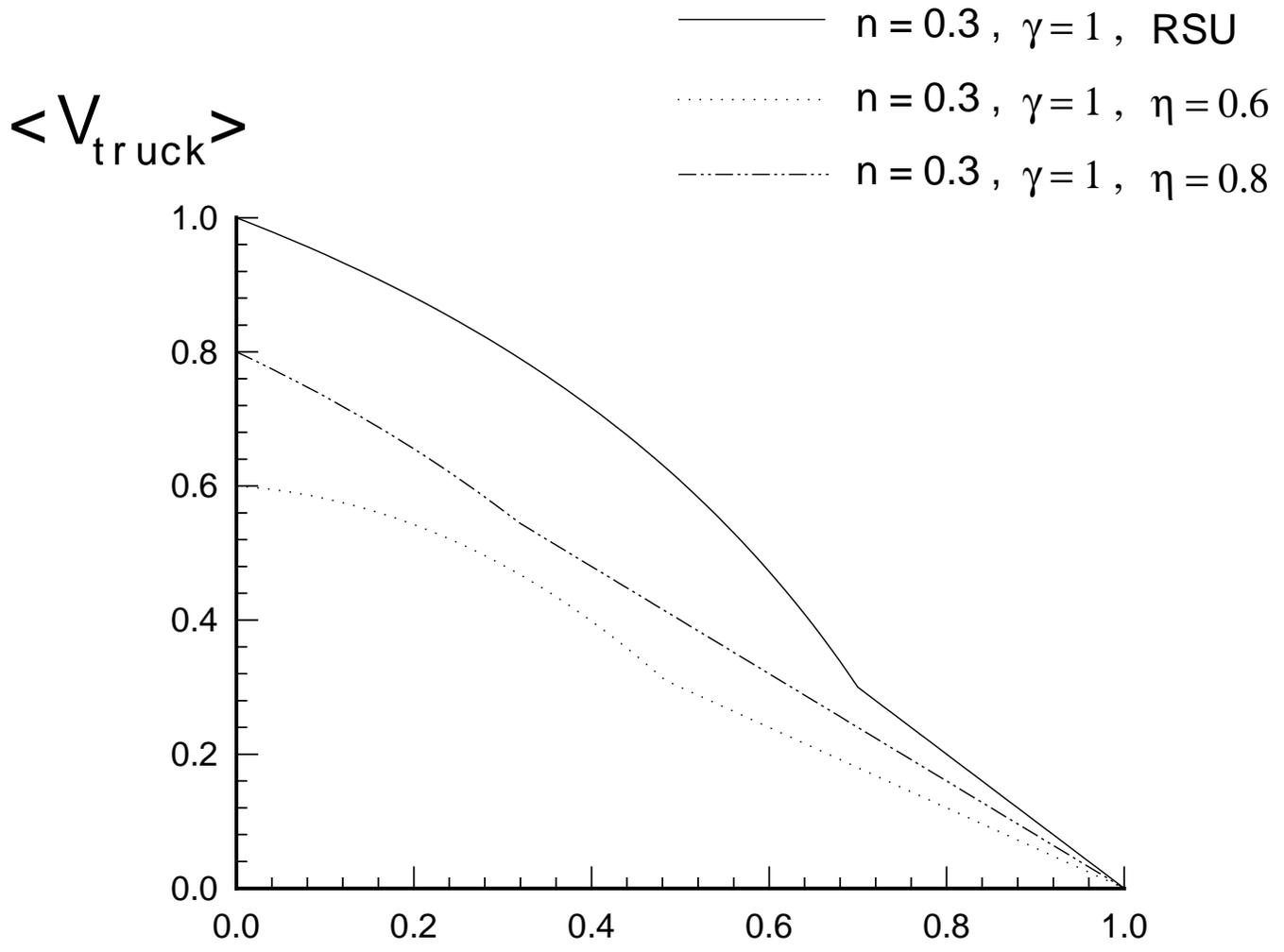
\begin{figure}
\caption{Average velocity of the truck in the FSU and RSU schemes for
$n=0.3$ and different values of $\eta $ }
\end{figure}

\begin{figure}
\caption{Fundamental diagrams in three updating schemes for
different values of $\beta$ and $\eta$ (values are given
in the figure)}
\end{figure}

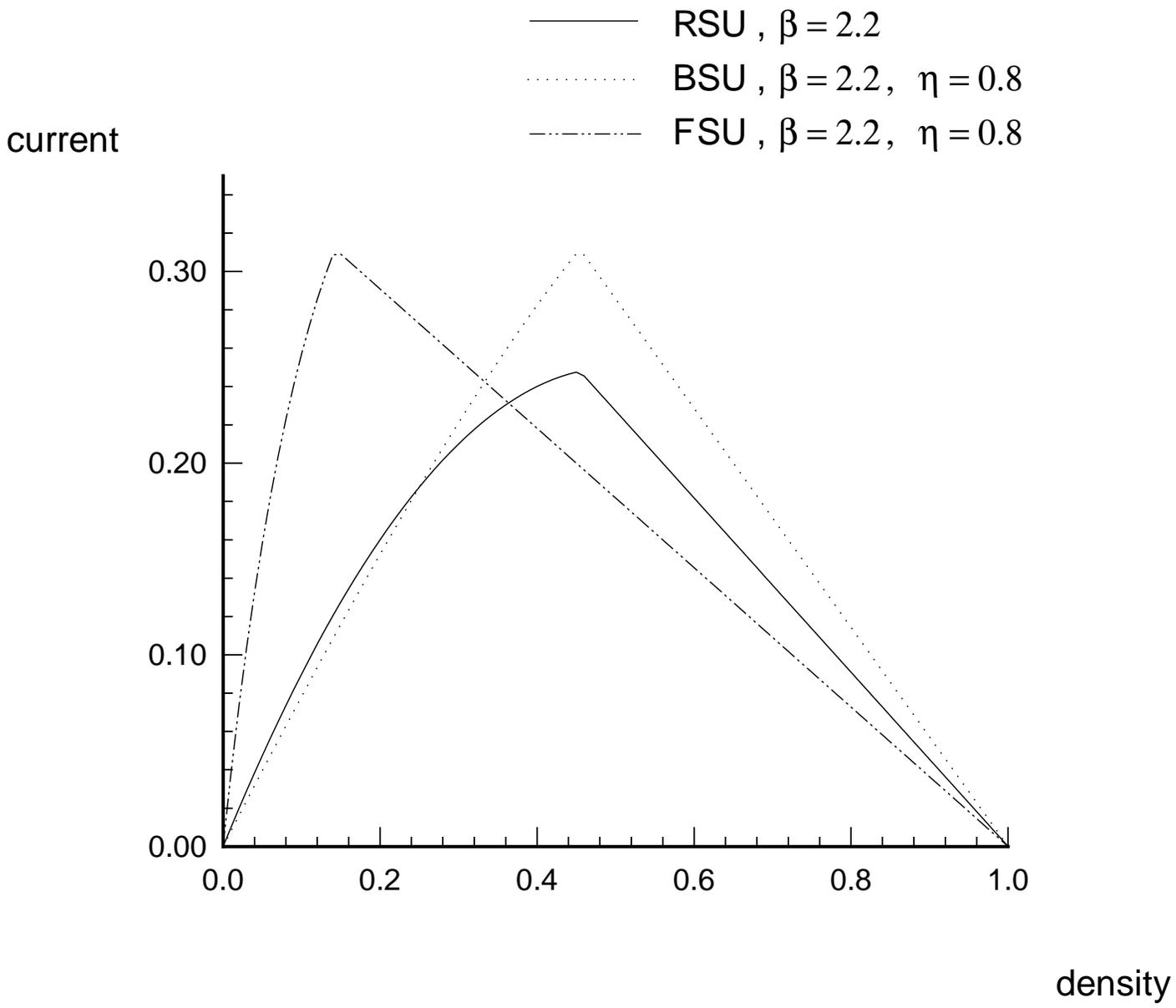
\end{document}